\documentclass{aa}
\usepackage{txfonts}
\usepackage{graphicx}
\usepackage{natbib}
\bibpunct{(}{)}{;}{a}{}{,}

\begin{document}

\title{Gravitational scattering by giant planets}

\author{T. Laakso\inst{1}
\and J. Rantala\inst{2}
\and M. Kaasalainen\inst{1}}

\institute{Department of Mathematics and Statistics, Rolf Nevanlinna
Institute, PO Box 68, 00014 University of Helsinki, Finland \and
Observatory, PO Box 14, 00014 University of Helsinki, Finland}

\date{Received 1 March 2006 / Accepted 1 June 2006}

\abstract{We seek to characterize giant-planet systems by their
gravitational scattering properties. We do this to a given system by
integrating it numerically along with a large number of hypothetical
small bodies that are initially in eccentric habitable zone
(HZ)-crossing orbits. Our analysis produces a single number, the
escape rate, which represents the rate at which the small-body flux is
perturbed away by the giant planets into orbits that no longer pose a
threat to terrestrial planets inside the HZ. Obtaining the escape rate
this way is similar to computing the largest Liapunov exponent as the
exponential rate of divergence of two nearby orbits. For a terrestrial
planet inside the HZ, the escape rate value quantifies the
``protective'' effect that the studied giant-planet system
offers. Therefore, escape rates could provide information on whether
certain giant-planet configurations produce a more desirable
environment for life than the others. We present some computed escape
rates on selected planetary systems, focusing on effects of varying
the masses and semi-major axes of the giant planets. In the case of
our Solar System we find rather surprisingly that Jupiter, in its
current orbit, may provide a minimal amount of protection to the
Earth.

 }

\maketitle


\section{Introduction}

The fascinating dynamical variety of extrasolar planetary systems has
been the motivation behind many recent numerical simulations. We have
seen detailed studies concerning the orbital stability of observed
multi-planet systems \citep[e.g.,][]{BQ2004} and also stability
analyses where fictitious terrestrial planets are integrated along with
the observed giant-planet system
\citep[e.g.,][]{JUS2005,MT2003,ABC2004}. In a few cases (GJ 777A, 47
UMa, HD 4208, HD 72659 etc.) it is found that terrestrial planets
could indeed survive inside the habitable zone (HZ) of the system in a
million year time scale.

For those candidate systems that could harbor Earth-like planets, we
can take our dynamical speculation one step further and ask: How
intense would the small-body flux be in the habitable zone of the
planetary system in question? After all, small-body impacts to the
Earth have had a major influence on the evolution of terrestrial
life. Although it is arguable whether a smaller or larger small-body
flux would have been more beneficial for our evolution, it would still
be a step forward if we could present something quantitative about the
small-body fluxes in the extrasolar systems.

The initial conditions in the circumstellar disk, the process of
planet formation, and subsequent dynamical evolution undoubtly sculpt
an individual distribution of small bodies for each planetary system.
A recent analysis of the debris disk around $\tau$ Ceti
\citep{GWH2004} suggests that the disk has similar dimensions, but
mass an order of magnitude greater than the Kuiper Belt, indicating a
more intense flux of cometary bodies. Is seems that, again, our solar
system is not necessarily to be taken as the prevalent specimen, and
that the bombardment by small bodies may become an important variable
when searching for habitable Earth-like planets. Direct observations
of debris disks are currently limited to very nearby stars, and are
unavailable for the majority of extrasolar planetary systems found by
radial velocity measurements ({\it http://www.obspm.fr/planets}). For
these systems, models of planet formation could be used to produce a
hypothetical distribution of small bodies, but this would lead us to
undesirable complex simulations and heavy computational load. Hence,
we accept the fact that their current population of small bodies is
unknown, and suggest another kind of method of analysis.

In our approach we follow the evolution of a specific hypothetical
population of small bodies which is common to all of the planetary
systems under study. Our purpose is to isolate and characterize the
small-body scattering properties of the planets, and to provide a
technique to compare different planetary systems in this respect. If
we can show that a certain configuration of planets is considerably
more efficient in scattering small bodies than another one, we could
argue that the actual small-body flux is less intense in that system,
taken that the initial conditions in both systems are
similar. \citet{Wet1994} used this idea when he considered alternative
giant-planet formation scenarios in our Solar System. Using
{\"O}pik-Arnold calculations he followed the evolution of cometary
test bodies and concluded that the absence of Jupiter would increase
the cratering rate on Earth throughout its history by a factor of
100--1000. The dynamically dominant role of Jupiter in the Solar
System was also shown by \citet{Eve1968}, in the case of parabolic
comets.

In this paper we will present computational tools to measure the
gravitational scattering properties of a planetary system. First, in
Sect.\ \ref{s:methods}, we will introduce a ``benchmark'' integration
scenario with thousands of cometary small bodies to be integrated
numerically along the planetary system under study. The small bodies
are initially in orbits that cross the HZ of the system. We will also
define a variable, the escape rate, that describes the strength at
which the planetary system scatters the small bodies into orbits that
no longer threaten the HZ. In Sect.\ \ref{s:results}, we will present
some preliminary results obtained using the tools above. We have
considered hypothetical giant-planet configurations based on the Solar
System and especially evaluated the dynamical significance of
Jupiter. Later, we will show some results on selected extrasolar
planetary systems. In Sect.\ \ref{s:discussion} we will discuss the
interpretation of the results.


\section{Methods}
\label{s:methods}

\subsection{The escape rate of particles}
\label{s:escape_rate}
We study planetary systems of one or more giant planets orbiting a
single Sun-like star. Into the planetary system we place a swarm of
massless test particles representing a population of HZ-threatening
small bodies. The initial orbital distribution of the particles is
random, but chosen such that the pericenter distances are inside the
HZ. As the system is propagated in time by numerical integration, the
orbits of the particles are perturbed by the giant planets, deviating
some of the particles into orbits that are no longer
HZ-threatening. Our assumption is that this gravitational scattering
is characteristic to the giant-planet system, and that its strength
can be measured straightforwardly, as follows.

Suppose that in the scenario described above, we are in a state where
the probability of a particle to be scattered per unit time, $\gamma$,
is constant. This implies that at any time $t$ the size of the
particle population $N$ changes by $dN=-N\gamma dt$, and that the
population depletes exponentially;

\begin{equation}  
N\left(t\right)=N_0\mathrm{e}^{-\gamma\left(t-t_0\right)},
\label{e:depletion}
\end{equation}

\noindent where $N_0$ is the initial population at time $t_0$. We
could try to determine $\gamma$ simply by integrating the system from
$t_0$ to $t$ and using (\ref{e:depletion}), but some difficulties
would arise: first, a suitable choice for $N_0$ and $t-t_0$ would
depend on $\gamma$, i.e., on the scattering properties of the
planetary system. Second, in reality, it is unlikely that we would see
exactly exponential depletion of particles with small values of
$N$. Because of these issues, we define $\rho$, an approximation to
$\gamma$, similarly to the method of computing Liapunov characteristic
exponents (LCEs) \citep[see, e.g.,][]{LL1992}, as

\begin{equation} 
\rho=\frac{1}{\left(t-t_0\right)}
\left(\sum_{i=1}^{k}\ln\frac{N_0}{N\left(t_i\right)}+
\ln\frac{N_0}{N\left(t\right)}\right), 
\label{e:escape_rate} 
\end{equation} 

\noindent where $t_0<t_1<...<t_k<t$. At the intermediate time points
$t_1,...,t_k$ the population is renormalized (regenerated) back to its
initial level $N_0$ by creating new random particles from the initial
orbital distribution. In this way we can keep $N(t)$ always inside
predefined boundaries and select a common $N_0$ and $t-t_0$ for all
the planetary systems we investigate. If a system scatters particles
at a constant rate, we should see $\rho$ converging towards a specific
value. We refer to this value as the escape rate. There is a
resemblance to the escape-rate formalism in statistical mechanics
\citep[e.g.,][]{Dor1999}. We emphasize the analogy between the
computation of $\rho$ and LCEs. Instead of exponential divergence of
nearby orbits, we monitor the exponential depletion rate of particles
in a large population. The renormalization is done for similar reasons
in both cases; with $\rho$ its purpose is to exclude possible effects
caused by the population not being ``large'' anymore.

We would also like to point out that our intention is to define the
escape rate as a tool for statistical physics rather than as a
measurement for actual physical quantities. Numerically computed
escape rate values always depend on the initial orbital parameter
distribution of the particle swarm. Therefore, the escape rate tells
us how the planetary system responds to a certain type of initial
population. In a specific application, in order to improve the
informative value of the escape rate, the population could be chosen
to represent, e.g., HZ-crossing comets originating from an Oort-type
reservoir. In this paper, however, we choose an initial population
which makes the demonstration of our method as simple as possible, but
still allows us to make indicative arguments about real planetary
systems.

\subsection{Parameters for the particle swarm}
\label{s:parameters}
In this section we define the initial and critical orbital elements
for the particle swarm, and the dimensions of the HZ. These parameters
fix a benchmark integration scenario which we use for computing escape
rates in this paper.

Obtaining proper convergence in escape rate computation requires that
the flux of particles throughout the planetary system remains as close
to steady state as possible. If the initial orbital distribution that
feeds new particles into the system is far away from the steady-state
distribution, we see a secular drift in the escape rate value. In the
process of finding a suitable and simple orbital distribution we
learned that, in order to enhance the convergence of the escape rate
computation, we should:
\begin{itemize}
\item Choose an initial orbital distribution that is relatively
homogeneous and contains only chaotic orbits. In other words, orbits
that have strong interaction with the giant planets and a mutually
similar timescale for being scattered. 
\item Select an initial orbital distribution that is spherically
symmetric. A disk-shaped distribution is not close to steady state,
since many particles are perturbed into high inclination orbits before
being scattered.
\item Introduce an offset time to the integration, after which the
escape rate computation is started. This eliminates the effect of an
initial transient period when the flux of particles seeks its
equilibrium.
\end{itemize}
Following these guidelines, we define the initial particle population
for our benchmark integration; each orbital element is randomly
selected from an even distribution bounded by the intervals in Table
\ref{t:elements}.
\begin{table}[hbt]
\caption{Initial orbital elements for the particle swarm.}
\label{t:elements}
\centering
\begin{tabular}{lcl}
\hline\hline
orbital element & & interval \\
\hline
pericenter distance & $q$ & $[0.5, 1.5]$ AU \\
apocenter distance & $Q$ & $[10, 80]$ AU \\
inclination & $\iota$ & $[0,\pi]$ \\ 
longitude of the ascending node & $\Omega$ & $[0,2\pi]$ \\
longitude of perihelion & $\tilde{\omega}$ & $[0,2\pi]$ \\
mean longitude & $L$ & $[0,2\pi]$ \\
\hline
\end{tabular}
\end{table}
The interval for pericenter distance $q$ coincides with our definition
for the inner ($0.5$ AU) and outer ($1.5$ AU) edges of the HZ. This is
a rough approximation; a detailed analysis of HZs around main sequence
stars is given by \citet{KWR1993}. We adopt such a broad range of
values since we want it to cover the actual HZs of most of the
observed extrasolar systems. In addition, by using a common definition
for the HZ we can compare the escape rates of systems with different
central stars.

During the integration, particles scattered into orbits that will no
longer cross the HZ should contribute to the escape rate value and be
removed. We define and monitor the following alternative criteria for
particle removal:
\begin{itemize}
\item Pericenter distance $q>2.0$ AU.
\item Ejection out of the system (semi-major axis $a>10000$ AU or
hyperbolic orbit).
\item Collision with one of the planets or the central star.
\end{itemize}

\subsection{Orbit integration}
The escape rate computation requires a numerical integration method
that has the following properties:
\begin{itemize}
\item The integrator must be able to handle close encounters between
planets and massless bodies.
\item High numerical accuracy is not important, but the method must be
robust.
\item The method has to be fast and optimized for integrating a
planetary system with thousands of massless bodies.
\end{itemize}
We use a hybrid integration scheme where the primary propagator is the
mixed variable symplectic (MVS) leapfrog method \citep[see,
e.g.,][]{MD1999}. Close encounters are handled with the standard
Bulirsch-Stoer algorithm \citep{SB1992} using the full force
function. Our integration method is close to the method presented by
\citet{Cha1999}, the main difference being that we use heliocentric
and Jacobian coordinates, and have not implemented a changeover
function for more sophisticated switching between the integration
methods. Since our close encounters are always between a massive and a
massless body, the Hamiltonian is trivially conserved. Table
\ref{t:integration} lists the integration parameters we use in the
benchmark integration.
\begin{table}[hbt]
\caption{Integration parameters in benchmark runs.}
\label{t:integration}
\centering
\begin{tabular}{lll} 
\hline\hline
number of particles & $N_0$ & $10\ 000$ \\ 
integration time & $t$ & $100\ 000$ years \\ 
escape rate offset time & $t_0$ & $50\ 000$ years \\ 
leapfrog step size & $\tau_{lf}$ & $\sim 40$ days \\ 
Bulirsch-Stoer step size & $\tau_{BS}$ & $\sim 0.05-5$ days \\
\hline 
\end{tabular} 
\end{table}
 
We have implemented the integrator in standard Fortran 95 using
object-oriented programming techniques. It is part of a larger
numerical software package that we intend to make public in the
future. The Finnish Center of Scientific Computing (CSC) provides us
computing time on their IBM eServer Cluster 1600, a supercomputer with
512 Power4-processors. A benchmark run on a single processor takes
$6-8$ hours of CPU-time.

Since the massless particles do not interact with each other, the
integration algorithm can be parallelized by distributing the
particles evenly between the available processing units. In test runs,
we have obtained a speedup factor of one magnitude by using 16 CPUs
instead of one. However, with our current parameters, it is more
efficient to run multiple benchmarks simultaneously, each with a
single processor.


\section{Results}
\label{s:results}

\subsection{Integration consistency}
Before the scientific runs, we analysed the consistency of our
integrator. We used two different test setups in order to determine
the integration errors separately for planets and massless
particles. In the first test, we integrated the four giants planets of
the Solar System for one million years (the step sizes were set
according to Table \ref{t:integration}). This was effectively a test
for the MVS-part of the integrator, since there are no mutual close
encounters between the planets. The relative energy error of our
implementation shows similar behaviour to previously published ones
\citep[e.g.,][]{Cha1999}, that is, it hovers around $10^{-7}$
throughout the integration time.

For the second test, we set up a system with the Sun, Jupiter, and 100
massless particles corresponding to the restricted three-body
problem. Otherwise, the parameters were the same as in the benchmark
runs (Table \ref{t:integration}). We monitored the Jacobi constants of
the particles and found that the average error was $7\times10^{-6}$
and the maximum $2\times10^{-4}$. By reducing all the step sizes by a
factor of ten, the average and maximum errors changed to
$5\times10^{-9}$ and $2\times10^{-7}$, respectively. For performance
reasons, however, we decided that the step sizes in Table
\ref{t:integration} are sufficient for our qualitative analysis.

\subsection{The role of Jupiter}
\begin{figure}[ht]
\resizebox{\hsize}{!}{\includegraphics{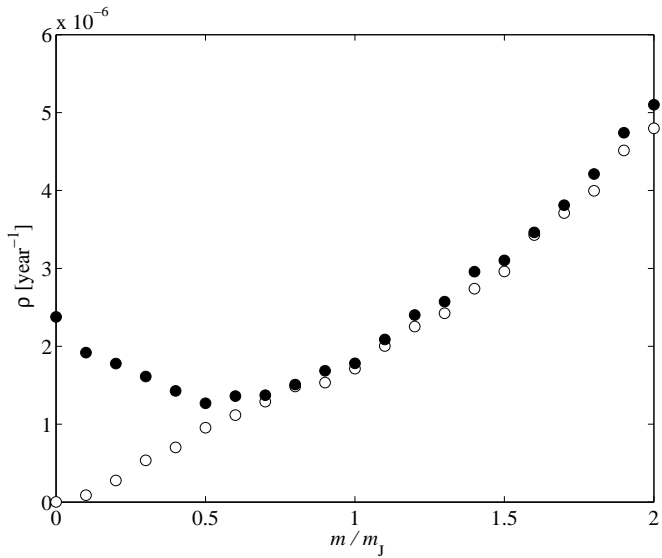}} 
\caption{Escape rates when Jupiter's mass is varied. System of
Jupiter, Saturn, Uranus, and Neptune is shown with filled circles, and
system of Jupiter alone with empty circles. $m_\mathrm{J}$ is the true
mass of Jupiter.}
\label{f:jmass} 
\end{figure}

We have computed escape rate values in order to study the dynamical
significance of Jupiter in the Solar System. We used two giant-planet
configurations; one with all of the giants (Jupiter, Saturn, Uranus,
and Neptune), and the other with Jupiter alone. The initial values for
the planets were given at J2000 epoch. In both configurations, we
varied the mass of Jupiter while retaining all other parameters. The
computed escape rate values are plotted in Fig.\ \ref{f:jmass}.

\begin{figure}[ht]
\resizebox{\hsize}{!}{\includegraphics{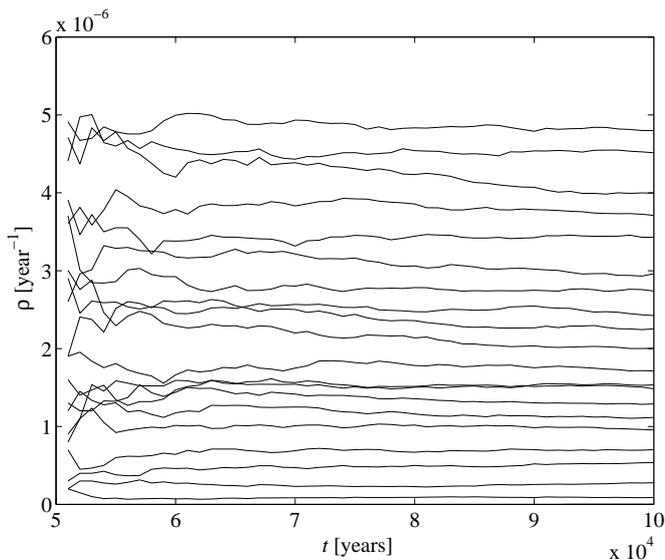}} 
\caption{Convergence of the escape rate values in integrations with 
Jupiter alone (cf. Fig.\ \ref{f:jmass}).} 
\label{f:convergence} 
\end{figure}

In the single planet case, we see close to linear dependence between
the mass of Jupiter and the escape rate. It is not surprising to
notice that a more massive planet scatters particles more efficiently,
but the responsiveness of the escape rate value encourages us to
believe that it could indeed possess some informative value. The
convergence of each escape rate computation in the single planet case
can be can be seen from Fig.\ \ref{f:convergence}. It seems that our
choices for the integration time and the number of particles are
sufficient, at least in this case.

Returning to Fig.\ \ref{f:jmass}, to the case where all four giant
planets are present, we now see more details in the escape rate. When
the mass of Jupiter is larger than its true value $1.0m_{\mathrm{J}}$,
the escape rate behaves similarly to the system with Jupiter
alone. This can be interpreted as Jupiter being the dominant scatterer
of particles among the other giants. However, for Jupiter masses below
$1.0m_{\mathrm{J}}$, the four-giant case eventually separates from the
Jupiter-only case, and the other giants take over in particle
scattering, effectively replacing Jupiter as its mass goes to
zero. There is a minimum in the escape rate value which roughly
coincides with the point of separation. We do not have a proper
explanation to this phenomenon, but one should consider the following
detail: the masses of Saturn ($m_{\mathrm{S}}$), Uranus
($m_{\mathrm{U}}$), and Neptune ($m_{\mathrm{N}}$) satisfy
$m_{\mathrm{S}}+m_{\mathrm{U}}+m_{\mathrm{N}}\approx
0.4m_{\mathrm{J}}$. Hence, when the mass of Jupiter is
$0.6m_{\mathrm{J}}$ the total mass of the four giants is approximately
$1.0m_{\mathrm{J}}$. We cannot confirm that this value is somehow
special for the system, but interestingly, it approximately coincides
with the point of separation and the minimum.

\begin{figure}
\resizebox{\hsize}{!}{\includegraphics{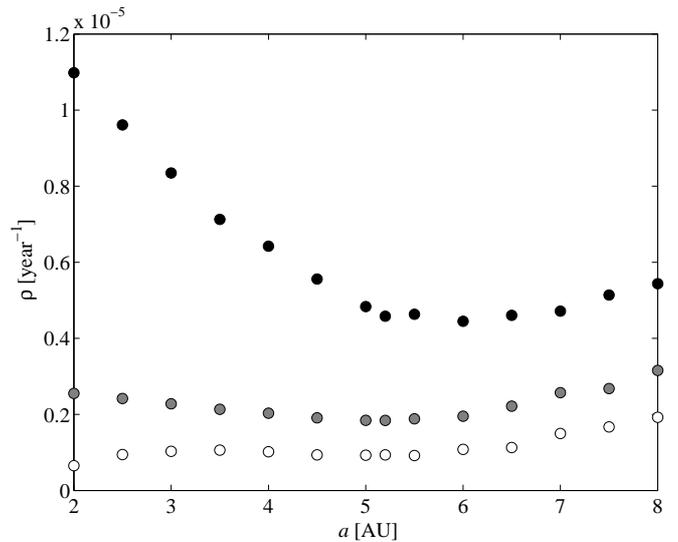}}
\caption{Escape rates for the Jupiter-only case when the semi-major
axis $a$ of the planet is varied. The mass of Jupiter is $2.0\
m_{\mathrm{J}}$ (black circles), $1.0\ m_{\mathrm{J}}$ (gray circles),
and $0.5\ m_{\mathrm{J}}$ (empty circles).}
\label{f:jsemi3} 
\end{figure}

Besides the mass, we were interested how changes in the semi-major
axis of Jupiter would affect the escape rate. We took the system with
Jupiter alone with three different masses ($0.5m_{\mathrm{J}}$,
$1.0m_{\mathrm{J}}$, $2.0m_{\mathrm{J}}$) and varied the semi-major
axis of the planet (Fig.\ \ref{f:jsemi3}).

\begin{figure} 
\resizebox{\hsize}{!}{\includegraphics{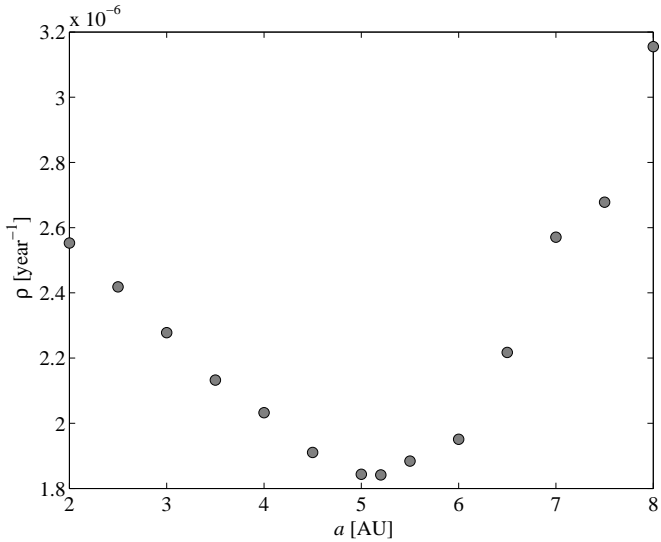}} 
\caption{Escape rates for the Jupiter-only case ($1.0\
m_{\mathrm{J}}$, isolated from Fig.\ \ref{f:jsemi3}) when the
semi-major axis $a$ of the planet is varied.}
\label{f:jsemi1}
\end{figure}

It seems that increasing the mass of the planet has greater impact on
the scattering strength than varying its semi-major axis. At least
with larger masses of Jupiter ($1.0m_{\mathrm{J}}$ and
$2.0m_{\mathrm{J}}$), we can identify a minimum in the escape rate
occurring at a certain value of the semi-major axis. The case where
Jupiter has its true mass is isolated in Fig.\ \ref{f:jsemi1}. It is
interesting to notice that the minimum occurs approximately at 5.2 AU,
at the true semi-major axis of Jupiter.

By monitoring the orbital parameters of the scattered particles we can
identify two trends that, when combined, explain the
minimum. Remembering our criteria for removing particles from
integration, we see that a Jupiter with $a>5.2$ AU is increasingly
effective at pulling the pericenter distances of particles above the
$2.0$ AU limit. On the other hand, the closer a giant planet is to the
HZ the stronger are the perturbations on particles near their
pericenters, resulting in numerous ejection orbits.

\subsection{Extrasolar systems}
We computed the escape rate values for two known extrasolar planetary
systems; \object{47 UMa} and \object{GJ 777A} (\object{HD
190360}). Table \ref{t:exoplanets} shows our initial parameters for
the planets. The data for \object{47 UMa} is given by
\citet{FMB2002}. The system is one of the best candidates for having
earth-like planets. Until recently, \object{GJ 777A} was also
considered as a feasible candidate. However, measurements by
\citet{VBM2005} may change this picture, since they suggest that there
is a second planet orbiting close to the central star. Nevertheless,
we chose to analyse \object{GJ 777A}, because it now represents
a qualitatively different case where the HZ lies between the orbits
of two perturbing planets.

\begin{table}[hbt]
\caption{Initial parameters for the extrasolar planets.}
\label{t:exoplanets}
\centering
\begin{tabular}{l|ll|ll}
\hline\hline
parameter & 47 UMa b & 47 UMa c & GJ 777A c & GJ 777A b\\
\hline
$a$ [AU] & $2.09$ & $3.73$ & $0.128$ & $3.92$\\
$e$ & $0.061$ & $0.1$ & $0.01$ & $0.36$\\
$\tilde{\omega}$ [$^\circ$] & $172.0$ & $127.0$ & $153.7$ & $12.4$\\
\hline
$m$ [$m_J$] & $2.54$ & $0.76$ & $0.057$ & $1.502$\\
\hline
\end{tabular}
\end{table}

Inclinations $\iota$, longitudes of the ascending node $\Omega$, and
mean longitudes $L$ were all set to zero. Masses of the central stars
were $1.03$ $m_{\sun}$ for \object{47 UMa} and $0.96$ $m_{\sun}$ for
\object{GJ 777A}. The computed escape rate values are shown in Table
\ref{t:exoresults}, and compared to the Solar System (with four
giants).

\begin{table}[hbt]
\caption{Escape rates for selected extrasolar systems.}
\label{t:exoresults}
\centering
\begin{tabular}{lc} 
\hline\hline 
planetary system & $\rho$ [year$^{-1}$]\\ 
\hline
Solar System & $1.78\times 10^{-6}$ \\
47 UMa & $2.08\times 10^{-5}$ \\
GJ 777A & $4.25\times 10^{-5}$ \\
\hline 
\end{tabular} 
\end{table}


\section{Discussion}
\label{s:discussion}

We have introduced a technique for measuring gravitational scattering
efficiency in a planetary system. One simple dynamical quantity, the
escape rate, describes the protection that a giant-planet system
offers to a hypothetical terrestrial planet. Our results show that, at
least with our current choice of parameters, the escape rate is a
computationally consistent quantity and is a function of the orbital
structure of the giant planets. The important question is: can we use
escape rates to make conclusions that apply to real planetary systems
or are they just a dynamical curiosity with no physical significance?

We believe that the very definition for the escape rate (Sect.\
\ref{s:escape_rate}) is rigorous and, when comparing planetary
systems, the one with a higher escape rate would be more efficient in
scattering small bodies away from the HZ. However, this argument
applies only to the particular initial population of small bodies used
in the escape rate scenario. Hence, it is the choice of the population
that mostly determines what the escape rate really tells us.

There is an obvious conceptual difference that should be noted. In
reality, the shape and intensity of the small-body flux in a planetary
system is a function of time, whereas in the escape rate computation,
the flux is intentionally kept constant. Therefore, each escape rate
scenario represents a fixed moment in time. In order to simulate the
small-body flux in a real planetary system, we should fix a time in
the history (or in the future) of the system, and create an initial
population that is a model for the HZ-crossing flux at that particular
moment. Unfortunately, by doing so, we would inevitably lose the
ability to compare the escape rates of totally different planetary
systems. This kind of approach could still be valid and beneficial, if
we concentrated to a one well known system (i.e., to our Solar System)
and varied the parameters of the planets only by small amounts.

In the benchmark integration scenario presented in this paper (Sect.\
\ref{s:parameters}) we use the escape rate as a universal dynamical
quantity which is not identified with real fluxes of small bodies. Our
initial population is artificial, but for the sake of creditability,
it should still be somehow ``typical'' among planetary systems.  Our
distribution of particles is spherically symmetric, and in the orbital
parameter space the particles cover the angular dimensions ($\iota$,
$\Omega$, $\tilde{\omega}$, $L$) completely, but only a limited range
in semi-major axis and eccentricity ($a$, $e$). This kind of
distribution is not typical to the observed small-body populations or
to the simulations explaining their evolution \citep[see, e.g.,][ for
Solar System comets]{DQT1987,DQT1988}. However, one should remember
that the initial population for the escape rate is supposed to
represent only the fraction of small bodies that threaten the HZ of
the planetary system. The modelling of all the processes that bring
small bodies into to HZ-crossing flux would be a cumbersome task and,
hence, we feel justified to use our simplified initial population as a
first order approximation.

As a part in improving the convergence and efficiency in escape rate
computations, we chose the initial orbital distribution in a way that
the apocenter distances ($Q$) were all well beyond the semi-major axis
of the innermost giant planet. Obviously, a qualitatively different
choice, e.g., one with the apocenter distances inside the orbits of
the giants, could change our results (Sect.\ \ref{s:results})
significantly. This is something we wish to investigate in future
papers. However, as an indication of partial robustness, the
qualitative behaviour of our results did not change when we used an
initial population with $Q=\left[40,50\right]$ and a narrower
definition for the HZ and for the removal criterion by particle
pericenter ($\left[0.8,1.4\right]$ and $1.5$ AU, respectively). Our
choice for the initial population also ensures that the orbits of the
particles are chaotic and none of the particles survive the whole
integration time. If the population included numerous stable or
resonant orbits, the regenerated particles might accumulate into safe
areas in the phase space, and the escape rate might become biased.

There are open questions about the initial population, but our results
with the benchmark scenario show that the escape rate has
informational value, and it can be used in comparative analyses on
planetary systems which was our goal in the first place. As such, the
escape rate cannot predict the habitability of a hypothetical
terrestrial planet in an observed extrasolar planetary system, but it
could be used, among other tools, to make educated guesses about the
matter.

The computed escape rate values for the Solar System giant planets
confirm that Jupiter has the dominant role in scattering small bodies
out of the HZ. The scattering efficiency in the Solar System would not
be greatly disturbed if the other giant planets were removed. This is
in line with the findings by \citet{Wet1994}. On the other hand,
removing Jupiter does not affect the escape rate considerably either
since the other giants effectively take its role in scattering.

The minimum in the escape rate at the true semi-major axis of Jupiter
is an interesting detail. It could be a coincidence, but perhaps there
is some cosmogonical, or dynamical, explanation. Maybe Jupiter, the
dominant mass, is in an optimal orbit to encourage the heavy
bombardment of cometary bodies into the inner Solar System. This could
have been an essential requirement for life and evolution on Earth.

If we compare the escape rate for the Solar System to the escape rates
for \object{47 UMa} and \object{GJ 777A} we see that small-body
scattering efficiency is more than an order of magnitude greater in
the extrasolar systems. A natural explanation to this are the smaller
semi-major axes and greater masses of the extrasolar giant
planets. Comparison between \object{47 UMa} and \object{GJ 777A} is
more interesting, because the higher escape rate of \object{GJ 777A}
is difficult to explain by intuition. By looking at the statistics
produced by the integration, we saw that the outer planet in
\object{GJ 777A} was the most recent perturber for $92\%$ of the
scattered particles. Therefore, the large eccentricity of the outer
planet is probably responsible for the higher escape rate value.

This is a preliminary paper where we have introduced a new concept,
implemented a computational method, and demonstrated some basic
applications. There are many possibilities for follow-up studies; an
important one is to analyse further the effect of the initial
population on the escape rate. Another task, which was beyond the
CPU-time budget of this paper, would be to probe the parameter space
of the Solar System giants more thoroughly. In addition to semi-major
axes and planetary masses, one could include the eccentricities into
the analysis. Generally, the increasing dimensionality of the
parameter space will certainly become a problem, but with some kind of
compromises, one could compute escape-rate maps which could reveal
interesting dynamical structures.


\begin{acknowledgements}
      This work was partly supported by the Finnish Cultural
      Foundation, Emil Aaltonen Foundation, Jenny and Antti Wihuri
      foundation, and Academy of Finland. We are grateful to an
      anonymous referee who provided us with thorough and constructive
      comments.
\end{acknowledgements}

\bibliographystyle{aa}
\bibliography{5121.bib}

\end{document}